\documentclass[amsmath,twocolumn,prb,aps]{revtex4-1}
\usepackage{amsthm,amsfonts,graphicx,verbatim, color}
\usepackage{bm}% bold math
\usepackage[T1]{fontenc}
\usepackage[utf8]{inputenc}
\usepackage{physics}
\usepackage[colorlinks=true, linkcolor = blue, citecolor = blue, urlcolor = blue]{hyperref}

\newcommand{\mb}[1]{\ensuremath{\mathbf #1}}

\newcommand{\aQH}{\ensuremath{\alpha_{\tiny \sf QH}}}
\newcommand{\laughlin}[1]{| \psi_L^{(#1)} \rangle}

\graphicspath{{./figs/}}

\begin{document}

%\begin{comment}
\title{Interaction-dependent anisotropy of fractional quantum Hall states}

\author{Akshay Krishna${}^1$, Fan Chen${}^1$, Matteo Ippoliti${}^2$, and R. N. Bhatt${}^1$}
\affiliation{${}^1$Department of Electrical Engineering and ${}^2$Department of Physics, Princeton University, Princeton NJ 08544, USA}

\begin{abstract}
A fractional quantum Hall (FQH) system with broken rotational symmetry exploits its geometric degree of freedom to minimize its ground state energy.
The mass anisotropy of bare particles interacting isotropically is partially inherited by the many-body FQH state, and the extent to which it does so depends on the type of interaction, filling fraction and ground state phase.
Using numerical infinite density matrix renormalization group simulations, we investigate the transference of elliptical ($C_2$-symmetric) anisotropy from the band mass of the bare particles to the FQH states, for various power law interactions. 
We map out the response of FQH states to small anisotropy as a function of power law exponent, filling, and statistics (bosonic or fermionic) of the constituents.
Interestingly, we find a non-analyticity in the linear response of the FQH state at a special filling-dependent value of the power law exponent, above which the interaction effectively becomes zero-range (point-like).
We also investigate the the effect of $C_4$-symmetric band distortions, where we observe a strikingly different dependence on filling.
%We provide an intuitive explanation of our findings using the ideas of the Laughlin wavefunction and Haldane pseudopotentials.
\end{abstract}

\maketitle

%%%%%%%%%%%%%%%%%%%%
%%%%%%%%%%%%%%%%%%%%
%%%%%%%%%%%%%%%%%%%%
%%%%%%%%%%%%%%%%%%%%

\section{Introduction \label{sec:intro}}

Fractional quantum Hall phases are extreme examples of strongly correlated matter.
In a high perpendicular magnetic field, the single-particle spectrum of a two-dimensional electron gas splits into highly degenerate Landau levels separated by cyclotron gaps which may be made arbitrarily large.
In this regime, a plethora of interesting phases can be realized \cite{Girvin1987, Yoshioka2013} as a function of Landau level filling, interaction type and disorder.
Some examples are 
(i) a gapped incompressible fractional quantum Hall (FQH) liquid~\cite{Tsui1982, Laughlin1983}, 
(ii) a gapless composite Fermi liquid (CFL)~\cite{Halperin1993}, 
and translation symmetry-breaking states such as 
(iii) Wigner crystals~\cite{Lam1984} and 
(iv) charge density waves (CDWs), including stripe and bubble phases~\cite{Moessner1996}.

Following Laughlin's variational wavefunction~\cite{Laughlin1983}, the hierarchy~\cite{Haldane1983} and flux attachment~\cite{Jain1989} pictures have paved the way for our understanding of many of the FQH plateaus.
%Laughlin's wavefunction perfectly describes the ground state of an FQH system with contact interactions but is not the true ground state for Coulomb interaction.
%However, for sufficiently short-ranged interactions, it has a high overlap with the true ground state and correctly explains a number of observed effects in the ground state such as the quantized Hall response, quasiparticle statistics etc.
This understanding was initially confined to rotationally symmetric Hamiltonians.
While it was known that the continuous rotational symmetry was not a necessary ingredient for FQH physics, and that it could be externally broken e.g. by an anisotropic band mass~\cite{Balagurov2000}, anisotropic FQH states received limited attention until Haldane~\cite{Haldane2011} pointed out the presence of an intrinsic geometric degree of freedom of FQH states, acting as a hidden variational parameter in the Laughlin wavefunction.
In the past decade, this geometric degree of freedom has received considerable attention~\cite{Qiu2012, Yang2012, Wang2012, Maciejko2013, Papic2013, You2014, Balram2016, Johri2016, Ciftja2017, Gromov2017A, Gromov2017B, Ippoliti2017, Ippoliti2017c, Ippoliti2017b, Yang2017, Yang2017B, Zhu2017,  Ippoliti2018, Lee2018, Liu2018, Zhu2018}.

Within this framework, theoretical efforts to understand the effects of anisotropy have focussed on anisotropic model wavefunctions \cite{Qiu2012, Balram2016} and pseudopotentials \cite{Yang2017, Yang2017B}.
These studies have been complemented by computational work on the effects of anisotropy due to band mass, interaction, tilted magnetic fields, and curved space~\cite{Yang2012, Wang2012, Papic2013, Johri2016}, as well as the role of filling fraction~\cite{Ippoliti2018}.
Connecting these results to experiments is not straightforward~\cite{Yang2016} as gapped states lack a Fermi contour and transport involves anisotropy of scattering as well.

In this work, we study the response of the ground state to anisotropy for various power-law interactions.
This provides us with a numerical probe of the `non-Laughlin-ness' of the true ground state as the interaction is tuned from Coulomb to shorter ranged.
We also investigate the response of FQH states on higher-order anisotropy, i.e. beyond the simplest case of elliptical distortion.
This case has not received as much attention, though a few studies exist \cite{Shen2015, Ciftja2004, Ippoliti2017b}. 

In contrast, for gapless fractions like $\nu = 1/2$, the presence of a composite Fermi contour makes direct experimental determination of the effects of Fermi surface deformation feasible \cite{Gokmen2010, Kamburov2013, Kamburov2014, Mueed2015, Jo2017, Jo2017b}.
In all cases, the competition between isotropic interaction and an anisotropic bare Fermi contour leads to a measurable effect on the anisotropy of the composite fermion Fermi surface.
Numerical calculations~\cite{Ippoliti2017} agree well with experimental observations; further the extent to which the anisotropy carries over to the composite Fermions depends on the interaction (e.g.\ its exponent for power-law interactions).
It is therefore of interest to see if the dependence of the CFL's response to anisotropy as a function of interaction is qualitatively similar or different to that of gapped FQH states.

Historically, the most popular experimental platforms for the FQH effect have been two-dimensional electron gases confined in semiconductor quantum wells~\cite{Tsui1982, Clark1988, Kukushkin1992, Pinczuk1993, Saminadayar1997} and, more recently, graphene~\cite{Bolotin2009, Dean2011, Feldman2012, Amet2015}.
In recent years, there have been efforts to synthesize the FQH states in non-electronic systems, for example using ultracold atoms~\cite{Regnault2003, Cooper2013, Yao2013} and photons~\cite{Schine2016, Schine2019}.
In these systems, inter-particle interaction is not expected to be Coulombic, and short range or contact interactions are typically assumed.
These are compelling motivations for us to study the interplay of anisotropy and interaction for both fermionic and bosonic quantum Hall fractions.

The effect of anisotropy of the Hamiltonian will depend on the Landau level in question.
%The effects of anisotropy on higher Landau levels are likely to be more complicated and interesting due to the proliferation of competing phases, and will be explored elsewhere.
Higher Landau levels are known to be more prone to instabilities towards rotational and translational symmetry breaking phases. The problem of anisotropy-induced phase transitions in $n>0$ Landau levels has received much attention both from theory \cite{Mulligan2011, Hu2018, Zhu2018} and from experiment \cite{Mueed2016, Shi2016, Hossain2018, Du2019} and holds many interesting open questions. 
In this paper, however, we focus on incompressible FQH states in the lowest ($n=0$) Landau level (LLL) that are stable to the application of anisotropy, and study their linear response to weak distortions that are far from any instability.
For strong enough distortions, we generally expect every FQH state to transition into a symmetry-broken phase. 
Characterizing this transition and its dependence on filling and interaction type is left to future work.

This paper is organized as follows. 
In Section \ref{sec:model}, we describe the Hamiltonian of our system and sketch the computational method underlying our calculations.
In Section \ref{sec:fermionic}, we provide a theoretical analysis of what should be expected in the case of fermionic FQH states, and present numerical results which are in agreement with those expectations.
Section \ref{sec:bosonic} describes our exploration of corresponding bosonic FQH states.
In Section \ref{sec:fourfold}, we present our results for $C_4$ symmetric distortions in the $\nu = 1/3$ and $1/5$ fermionic FQH states.
We conclude in Section \ref{sec:discussion} with a discussion of our results.

%%%%%%%%%%%%%%%%%%%%
%%%%%%%%%%%%%%%%%%%%
%%%%%%%%%%%%%%%%%%%%
%%%%%%%%%%%%%%%%%%%%

\section{Model and method \label{sec:model}}

Our system is described by the usual quantum Hall Hamiltonian for $N_e$ electrons in perpendicular magnetic field $\mb{B} = B \hat{\mb{z}}$, and corresponding magnetic vector potential $\mb{A}$:
\begin{align}
H &= \sum\limits_{i=1}^{N_e} T_i + H_{\text{int}} \nonumber \\
&= \frac{1}{2} \sum\limits_{i=1}^{N_e} (m^{-1})^{ab} \pi_{i, a} \pi_{i, b} + \frac{1}{2} \sum\limits_{i\neq j}^{N_e} V(\mb{r}_i - \mb{r}_j). \label{eq:hamiltonian}
\end{align}

The first term above is the kinetic energy, defined in terms of the dynamical momentum of the $i$\textsuperscript{th} electron $\bm{\pi}_i = \mb{p}_i - e \mb{A}_i$.
The inverse mass tensor is denoted $m^{-1}$.
Summation over the spatial indices $a, b \in \{ x, y\}$ is implicit.

The second term in Eq.\ \eqref{eq:hamiltonian} is the interaction potential $H_{\text{int}}$, which we take to be of the form $V(\mathbf r) \equiv V(|\mb r|_\varepsilon )$.
The distance $|\mb r |_\varepsilon$ depends on the dielectric tensor $\varepsilon$, which defines a spatially uniform metric %$\mathrm{d}r^2 = \varepsilon_{ab} \mathrm{d}r^a \mathrm{d}r^b$.
$|\mb r|^2_\varepsilon \equiv \varepsilon_{ab} r^a r^b$.
The two metrics described by $m_{ab}$ and $\varepsilon_{ab}$ are independent.
A linear change of coordinates can get rid of the anisotropy in either one of them, but not both simultaneously.
Without loss of generality, we take the interaction to be isotropic, $\varepsilon = \mathbf{1}$, which is accomplished by applying the linear transformation $\varepsilon^{-1/2}$ to the coordinates.
We can then rotate coordinates to make the mass tensor diagonal, though not proportional to the identity: $m_{ab} = \text{diag}(m_{xx}, m_{yy})$, $m_{xx} \neq m_{yy}$.
The single particle kinetic energy $T_i$ of Eq.\ \eqref{eq:hamiltonian} then becomes \begin{align}
T_i &=  \frac{\pi_{i,x}^2}{2 m_{xx}} + \frac{\pi_{i,y}^2}{2 m_{yy}}. \label{eq:singleparticleK}
\end{align}
We define $\alpha \equiv \sqrt{m_{yy} / m_{xx}}$ as the anisotropy of the non-interacting system.

We refer to this type of anisotropy, which is invariant under $C_2$ discrete rotational symmetry, as elliptical or two-fold anisotropy \cite{Ippoliti2017b}.
The first part of the present work analyzes this case.
In Sec.\ \ref{sec:fourfold}, we consider a generalization of the kinetic energy term of Eq.\ \eqref{eq:hamiltonian} to non-quadratic functions $T_i(\bm{\pi}_i)$ whose equal-energy contours are not ellipses.
Such a description allows us to study $N$-fold anisotropy ($N>2$), as may arise naturally from the symmetries of crystalline band structures.

In the limit of high magnetic field $B$, the cyclotron energy $\omega_c$ is much larger than other energy scales in the problem, and we may safely project all the dynamics to the lowest Landau level (LLL). 
In this limit, mixing with higher Landau levels is negligible.
The kinetic energy of the system is thus quenched, and the Hamiltonian reduces to \begin{align}
H_{\text{LLL}} = \frac{1}{2} \sum\limits_{i\neq j}^{N_e} \sum\limits_{\mb{q}} V(\mb{q}) |F_0(\mb{q})|^2 e^{i \mb{q} \cdot (\mb{R}_i - \mb{R}_j)}, \label{eq:hamiltonianLLL}
\end{align}
where $\mathbf{R}_i \equiv \mathbf{r}_i - \frac{l_B^2}{\hbar} \bm{\pi}_i \times \hat{\mb{z}}$ is the guiding center operator of the $i$\textsuperscript{th} electron and $V(\mb{q})$ is the Fourier transform of the interaction potential $V(\mb{r})$.
The form factor $F_0(\mb{q})$ accounts for the projection of the potential into the basis of anisotropic LLL orbitals, and encodes their anisotropy: 
\begin{align}
|F_0(\mb{q})|^2 = \exp \left[ - \frac{l_B^2}{2 } \left( \alpha q_x^2 + \frac{q_y^2}{\alpha} \right) \right]. \label{eq:formfactor}
\end{align}

We seek to compute the anisotropy of the many-body quantum Hall ground state, denoted by $\aQH$. This is facilitated by introducing a different parametrization of the anisotropies $\alpha$ and $\aQH$, following Ref.~\cite{Ippoliti2018}.
Since a $\pi/2$ rotation maps $\alpha \mapsto 1/\alpha$ and $\aQH \mapsto 1/\aQH$, in the 2D thermodynamic limit one has the reciprocity relation 
\begin{align}
\aQH(1 / \alpha) = \frac{1}{\aQH(\alpha)} \;. \label{eq:reciprocity}
\end{align}
In terms of the logarithmic quantities 
\begin{align}
\gamma \equiv \ln \alpha \quad \text{ and } \quad \sigma \equiv \ln \aQH \;,
\end{align}
Eq.~\eqref{eq:reciprocity} becomes
\begin{align}
\sigma(-\gamma) = -\sigma (\gamma) \;.
\end{align}
For small anisotropy, one can expand around the isotropic point $\gamma=\sigma=0$ and obtain
\begin{align}
\sigma (\gamma) \approx c_1 \gamma + \mathcal{O}(\gamma^3).
\end{align}
The quadratic coefficient is automatically absent due to symmetry, and the linear coefficient $0 \leq c_1 \leq 1$ quantifies the extent to which the band mass anisotropy is transferred to the quantum Hall state, with deviations $\mathcal O(\gamma^3)$.

For our numerical calculations, we set up the system on an infinite cylinder, with axis along $\hat{\mb{x}}$ and circumference $L_y$. Combined with the choice of Landau gauge $\mb{A} = Bx \hat{\mb{y}}$, this allows us to map the problem to a one-dimensional fermion chain and take advantage of the matrix-product states formalism.
We use the infinite Density Matrix Renormalization Group (iDMRG) algorithm for quantum Hall states~\cite{Zaletel2013, Zaletel2015}.
This introduces a second cut-off in the problem, besides $L_y$: the bond dimension $\chi$, i.e. the maximum dimension of the matrices used to approximate the many-body ground state.
After converging to the approximate ground state for a given anisotropy $\gamma$ and circumference $L_y$, we calculate the guiding center structure factor $S(\mb{q})$, defined below.
The effect of anisotropy is quantified by the long wavelength behavior of $S(\mb{q})$ (for gapped FQH states), or the non-analyticities of $S(\mb{q})$ at the composite fermion surface (for gapless CFL states).
We briefly review the procedure by which the anisotropy is calculated from $S(\mb{q})$ below.
A detailed discussion may be found in Ref.\ \cite{Ippoliti2018}. 

The guiding center structure factor $S(\mb{q})$ is defined as 
\begin{align}
%S(\mb{q}) = \frac{1}{N_\phi} \Bigg\langle \Big( \rho(\mb{q}) - \langle \rho(\mb{q}) \rangle \Big) \Big( \rho(-\mb{q}) - \langle \rho(-\mb{q}) \rangle \Big) \Bigg\rangle,
S(\mb{q}) = \frac{1}{N_\phi} \langle \delta \rho(\mb q) \delta \rho(-\mb q) \rangle,
\end{align}
where $\delta \rho(\mb q) \equiv \rho(\mb q) - \langle \rho(\mb q) \rangle$ and
\begin{align}
\rho(\mb{q}) = \sum\limits_{j=1}^{N_e} e^{i \mb{q} \cdot \mb{R}_j}
\end{align}
is the Fourier transform of the guiding center density operator.
Due to the incompressiblity of gapped FQH states, the structure factor is quartic at long wavelengths \cite{Girvin1985, Girvin1986}.
Since our simulations are on an infinite cylinder, we have access to a continuum of wavevectors $q_x$, and may express the long-wavelength structure factor\begin{align}
S(q_x, 0) \approx \lambda(\gamma) q_x^4 \qquad \text{as} \qquad q_x \to 0.
\end{align}
The prefactor $\lambda(\gamma)$ is obtained numerically by taking the limit \begin{align}
\lambda(\gamma) =  \lim\limits_{q_x \to 0} S(q_x, 0) /q_x^4 \; .
\end{align}
The dependence of this prefactor on $\gamma$ may be split into even and odd parts as $\lambda(\gamma) = e^{2(D(\gamma) + \sigma(\gamma))}$.
The even term $D(\gamma)$ represents an isotropic rescaling of the structure factor, and was found in Ref.~\cite{Ippoliti2018} to be nearly constant in Laughlin fractions with Coulomb interaction.% (with a very small term proportional to $\gamma^2$).
The term we are interested in is $\sigma(\gamma)$, which flips sign when we rotate the anisotropy by an angle of $\pi/2$ (i.e.\ transform $\gamma \to -\gamma$).
This is the term that controls the ellipticity of contours of $S(\mb{q})$ and is a useful proxy for the anisotropy of the many-body ground state wavefunction itself.
It is calculated as 
\begin{align}
\sigma(\gamma) = \frac{1}{4} \ln \frac{\lambda(\gamma)}{\lambda(-\gamma)} \;.
\end{align}
We sweep over a range of anisotropies $-0.3 < \gamma < 0.3$ ($0.74 < \alpha < 1.35$) to obtain a linear fit $\sigma(\gamma) \simeq c_1 \gamma$.

For the gapless state at $\nu = 1/2$, the structure factor $S(\mb{q})$ has singularities at $\mb{q}$ values corresponding to scattering processes between different points on the composite Fermi surface of the CFL.
%The locations of the singularities allow us to reconstruct the shape of the Fermi surface and extract the anisotropy of the ground state.
As described in Refs.\ \cite{Geraedts2016, Ippoliti2017}, placing the system on an infinite cylinder discretizes the momentum in the $\hat{\mb{y}}$ direction, $q_y \in (2 \pi / L_y)\mathbb Z$, so that the Fermi contour consists of isolated points.
Sharp features in $S(q_x, q_y)$ allow us to pinpoint the coordinates of these points. By aggregating data from multiple values of the circumference $L_y$, we can gather enough points to accurately reconstruct the elliptical shape of the two-dimensional Fermi contour and extract its anisotropy.

Capturing the behavior of the system in the thermodynamic limit from finite size iDMRG calculations requires care.
The circumference of the cylinder $L_y$ must be large enough to avoid spurious effects from periodic boundary conditions along that direction.
However, the entanglement entropy $S$ across a constant-$x$ cut in the cylinder obeys an area law, and thus grows linearly in $L_y$.
The bond dimension $\chi$ required to capture this entanglement accurately increases \emph{exponentially}: $\ln \chi \gtrsim S \sim L_y$.
Since the computational complexity of iDMRG is at least $\mathcal{O}(\chi^4)$, we are limited to a range of circumferences $14 l_B \leq L_y \leq 30 l_B$, for which bond dimensions $\chi \leq 4096$ provide accurate results.

The only form of the interaction $V(r)$ for which the analytical expression of $\sigma(\gamma)$ is known is that of a Gaussian interaction with characteristic length $s l_B$, i.e.\ $V(r) = e^{-\frac{1}{2}(r/s l_B)^2}$, due to Yang \cite{Yang2013}.
Since the form factor of the LLL is also a Gaussian, a clever rearrangement of terms provides a closed-form expression for the anisotropy of the quantum Hall state:
\begin{align}
\sigma(\gamma) &= \frac{1}{2} \ln \left( \frac{e^\gamma + s^2}{e^{-\gamma} + s^2}\right) \nonumber \\
&= \frac{1}{s^2+1} \gamma + \frac{s^2(s^2-1)}{6(s^2+1)^3} \gamma^3 + \cdots .\label{eq:GaussianYang}
\end{align}
The linear coefficient $c_1 = \frac{1}{s^2+1}$ decreases as the range $s$ of the isotropic interaction is made larger.

%At a fixed interaction length scale $s$, this should be true to all orders of $\gamma$ if the ground state of the isotropic system is stable.
%It should destabilize for large $s$.

For power law interactions, which we consider here, there is no simple expression for $c_1$, and we must compute it numerically as described in the following sections.

%%%%%%%%%%%%%%%%%%%%
%%%%%%%%%%%%%%%%%%%%
%%%%%%%%%%%%%%%%%%%%
%%%%%%%%%%%%%%%%%%%%

\section{Fermionic states \label{sec:fermionic}}

%The pseudopotential description provides great insight into the effect of interaction, which we describe below.
Any interaction $V(\mb{r})$ can be expanded in terms of its Haldane pseudopotentials $V_m$ in the lowest Landau level.
In terms of these pseudopotentials, the ground state energy $E_{\text{gs}}$ of a configuration of $N_e$ electrons with many-body wavefunction $\ket{\psi}$ is \begin{align}
E_{\text{gs}} &= \braket{\psi}{H_{\text{LLL}} | \psi} / \braket{\psi}{\psi} \nonumber \\
&= {N_e \choose 2} \sum\limits_{n=0}^{\infty} A_{2n+1} V_{2n+1}.
\label{eq:Vm_E}
\end{align}
Here $A_n \ge 0$ is the probability for two electrons to have relative angular momentum $n \hbar$, and $\sum_n A_n = 1$.
Antisymmetry ensures that any valid fermionic ground state wavefunction $\ket{\psi}$ has all even $A_n$ equal to zero.

The Laughlin wavefunction at filling $\nu = 1/m$ is 
\begin{align}
\psi_L^{(m)}(\mb{r}_1, \cdots, \mb{r}_{N_e}) = \prod\limits_{i > j}^{N_e} (z_i - z_j)^m \exp \left( -\sum\limits_{i=1}^{N_e} \frac{|z_i|^2}{4l_B^2} \right), \label{eq:Laughlin}
\end{align} 
where $z_j = (x_j - i y_j) / l_B$ is the complex valued position of the $j$\textsuperscript{th} electron.
This wavefunction has the additional feature that odd coefficients $A_n $ vanish for all $n < m$.
By Eq.\ \eqref{eq:Vm_E}, the Laughlin state $\laughlin{m}$ is a zero-energy ground state of any interaction with non-negative pseudopotentials such that $V_n = 0$ for all $n \ge m$; it is the highest-density ground state if all the $V_n$ with odd $n<m$ are non-zero.
%An example of such interaction is the contact repulsion $V(\mb{r}) = \nabla^2 \delta(\mb{r})$ (pure $V_1$).

For a power-law interaction $V(\mb{r}) = \frac{U}{(r/l_B)^p}$, where $U$ is the interaction energy scale, 
\begin{equation}
V_m = \left \{
\begin{aligned}
& \frac{\Gamma(m+1-\frac{p}{2})}{m! 2^p } U  & \text{ if } p<2(m+1) \\
& \infty & \text{ otherwise. }
\end{aligned}
\right.
\label{eq:Vm_powerlaw}
\end{equation}
The $m$\textsuperscript{th} Haldane pseudopotential $V_m$ is infinite when $p \geq 2(m+1)$, and diverges logarithmically as $p \to 2(m+1)$. 
This fact has implications for FQH states at different fillings as described below.

\subsection{Fermionic FQHE parent states at $\nu = 1/3$ and $\nu = 1/5$} \label{sec:fermionicparent}

%The FQH state at $\nu = 1/3$ is described by Laughlin state $\laughlin{3}$.

For a power-law interaction with exponent $p \ge 4$, the Haldane pseudopotential $V_1$ becomes divergent.
The Laughlin state $\laughlin{3}$, which has coefficient $A_1 = 0$, is then the only possible antisymmetric state with finite energy at $\nu = 1/3$.
Equivalently, if one normalizes the overall energy scale $U$ such that $V_1\equiv 1$, all the higher pseudopotentials vanish, and $\laughlin{3}$ is the only zero-energy state.
This implies that, for fermions in the LLL at filling $1/3$, any power law interaction with $p \ge 4$ becomes effectively a \emph{contact} interaction, $V(\mb r) \sim \nabla^2 \delta(\mb r)$.
Such an interaction does not introduce a metric $\varepsilon_{ab}$ into the problem, and the band mass tensor $m_{ab}$ alone sets the FQH state's geometry. 
In other words, the problem is isotropic up to a rescaling of the coordinates, hence the anisotropy of the bare fermions carries over completely to the many-body ground state, giving $\aQH = \alpha$. 
The linear response coefficient in particular is $c_1 = 1$.

If the interaction power-law exponent satisfies $p \ge 8$, then \emph{both} $V_1$ and $V_3$ diverge.
At $\nu = 1/3$, the Laughlin state $\laughlin{3}$ (which has $A_1 = 0$ but $A_3 \neq 0$) has a divergent energy $E \sim A_3 V_3$.
However it is still the ground state, as $V_1$ diverges more strongly than $V_3$.
If one regularizes the interaction by introducing a short-distance cutoff $\Delta$ (as we shall do later to ensure numerical stability), then as $\Delta \to 0$ one has $V_m \sim \Delta^{2(m+1)-p}$, hence $V_3/V_1 \sim \Delta^4$. 
Therefore the divergence of $V_1$ is dominant and  $\laughlin{3}$, being the unique state with $A_1=0$ at $\nu = 1/3$, remains the ground state.
For $\nu = 1/5$ instead the divergence of $V_3$ has an important effect: the Laughlin state $\laughlin{5}$ (with $A_1 = A_3 = 0$) becomes the unique finite-energy ground state, with the attendant conclusions about the transference of anisotropy.

\begin{figure}[ht!]
\centering
\includegraphics[width=\columnwidth]{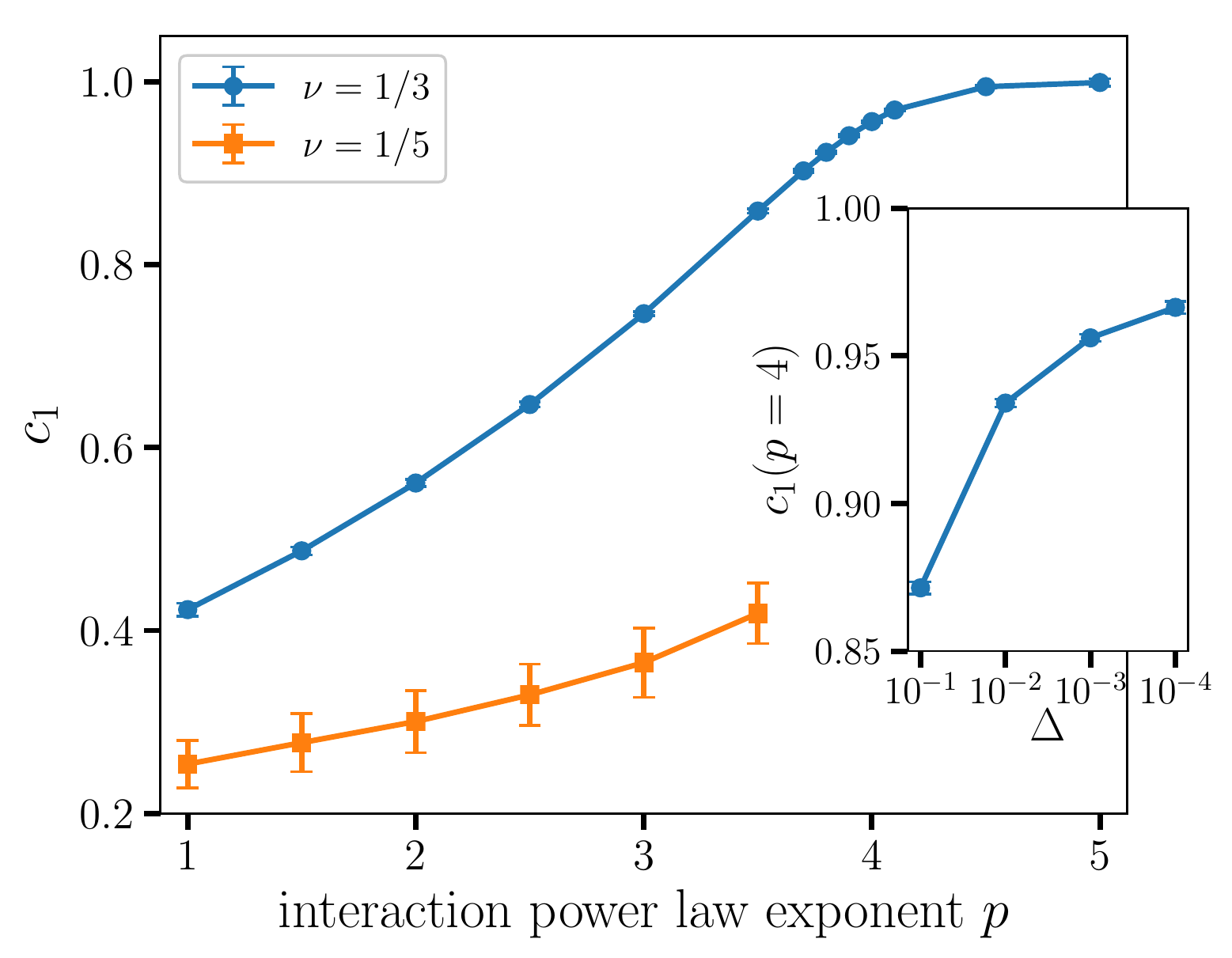}
\caption{The linear coefficient $c_1$ of the response of the $\nu=1/3$ and $\nu = 1/5$ FQH states to anisotropy is plotted as a function of interaction power-law exponent $p$. 
Simulations are carried out for 5 different sizes $L_y$ and bond dimensions $\chi = 2048$ to $4096$ to account for finite size and truncation effects.
We estimate errors from the standard deviation of the three best converged sizes at the highest bond dimension.
The short length-scale cut-off $\Delta$ is fixed at $10^{-3}$. 
(Inset) The value of $c_1$ at the critical power law $p_c=4$ for $\nu = 1/3$ is plotted as a function of the cutoff $\Delta$. 
It appears to flow to the theoretical value of $1$ as $\Delta \to 0$.}\label{fig:fermion_powerlaw}
\end{figure}

It is straightforward to see how this generalizes to different fractions:
for filling $\nu = 1/m$, there is a critical power law $p_c(m)  = 2(m-1)$ above which $\aQH=\alpha$ and thus $c_1 \equiv 1$. 
For $p<p_c(m)$, we instead expect the anisotropy of the bare fermions to carry over only incompletely to the quantum Hall state, giving $c_1 < 1$.

This picture implies a non-analytical behavior of $c_1$ as the interaction is made shorter ranged:
$c_1(p)$ cannot be analytical at $p = p_c$, as $c_1(p>p_c) \equiv 1$ while $c_1(p<p_c)$ is a nontrivial smooth function.
This may manifest as a kink (a discontinuity in the first derivative), a discontinuity in higher derivatives, or a more subtle non-analytical feature.
For comparison, in the case of Gaussian interaction $V(r) = e^{-r^2/2s^2l_B^2}$ (see Eq.\ \eqref{eq:GaussianYang}), we have $c_1 = (1+s^2)^{-1}$.
This is a smooth function of $s$ that flows asymptotically towards 1, in a \emph{filling-independent} manner, as the range is made shorter ($s \to 0$).

We numerically compute the anisotropy for a modified power-law interaction \begin{align}
V(r) = \frac{l_B}{r} \left( \frac{r^2}{l_B^2} + \Delta^2 \right)^{\frac{1-p}{2}} ,
\end{align} 
where $\Delta$ is a small regularizing parameter needed to ensure numerical stability at short length scales.
In the limit $\Delta \to 0$, we recover the familiar power-law with exponent $p$.
For Coulomb interaction ($p=1$), it was found~\cite{Ippoliti2018} that $c_1 \approx 0.43$.
The result of our numerical fit (Fig.\ \ref{fig:fermion_powerlaw}) for the linear anisotropy coefficient $c_1$ is consistent with $c_1$ increasing monotonically with the power-law exponent $p$ for $p<4$, and attaining a constant value $c_1=1$ for $p \ge 4$.
The deviation from the theoretical prediction near $p=4$ is entirely due to our use of a short range numerical cut-off $\Delta$.
As we reduce $\Delta$, the numerically obtained $c_1$ converges to the expected value.
As the singularity at $p=4$ is logarithmic, we must span several orders of magnitude in $\Delta$ to observe a significant drift of the result.
The approach to $\Delta \to 0$ is ultimately limited by numerical instability.
For the $\nu = 1/5$ FQH state, the value of $c_1$ is smaller than that for $\nu = 1/3$ at every power-law considered, indicating that the transference of anisotropy is much less in this case.
The value of $c_1$ also increases much more slowly as a function of power-law exponent, and is consistent with reaching a value of $c_1 = 1$ at $p=8$, although numerical stability issues limit the range of our study to smaller values of $p$.

\begin{figure}[ht!]
\centering
\includegraphics[width=\columnwidth]{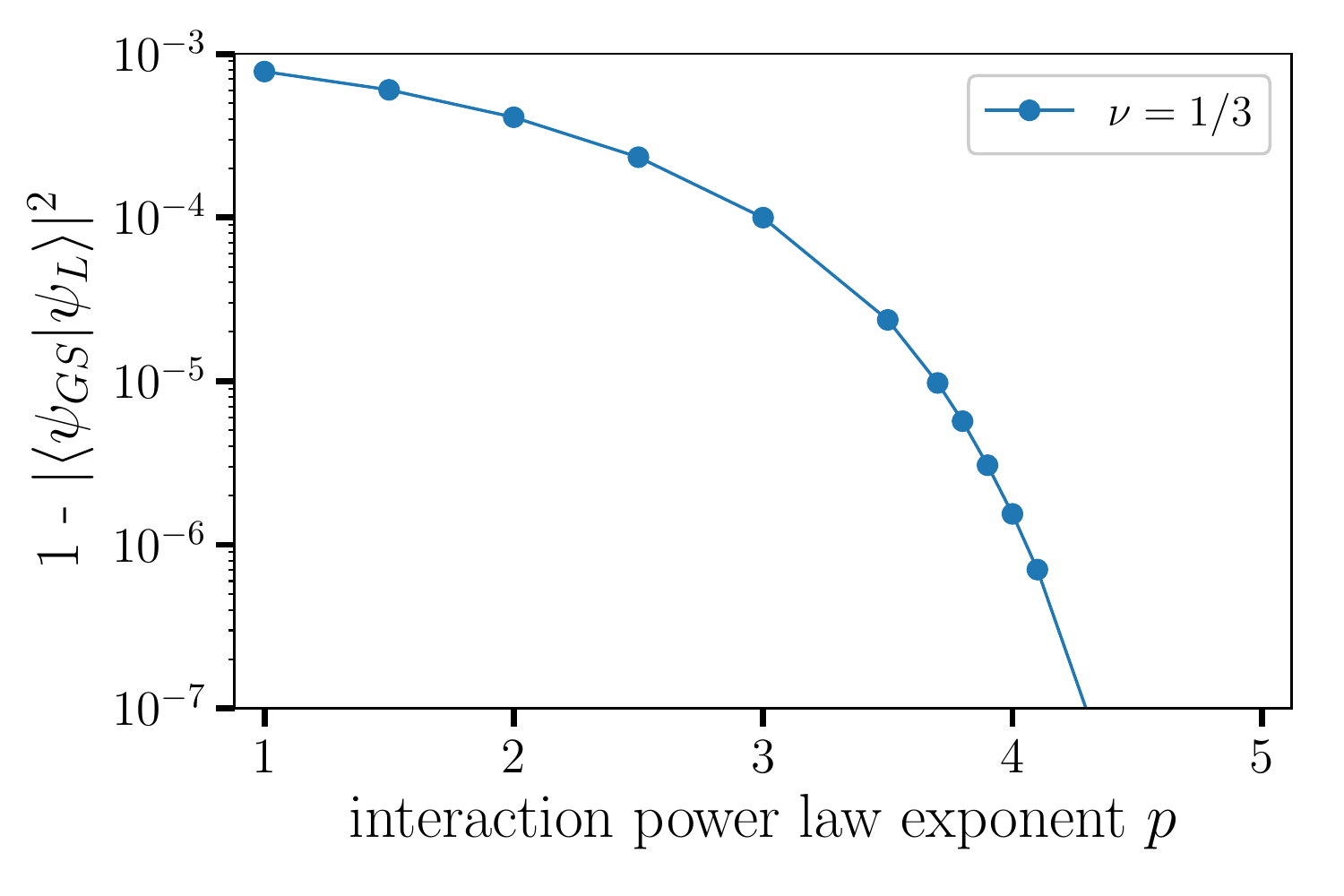}
\caption{The overlap (per quantum of flux) of the ground state of the power-law interaction at $\nu = 1/3$ with that of the Laughlin wavefunction $\laughlin{3}$ is plotted as a function of the power-law exponent at zero anisotropy. The system size is fixed at $L_y = 20 l_B$, the cut-off $\Delta = 10^{-3}$ and bond dimension for iDMRG is $\chi = 4096$.}\label{fig:fermion_laughlinoverlap}
\end{figure}

In Fig.\ \ref{fig:fermion_laughlinoverlap}, we show the overlap (per flux quantum) of the ground state of the power-law interaction, as obtained by the iDMRG algorithm, with that of the Laughlin wavefunction, obtained for a $V_1$ Haldane pseudopotential interaction.
The overlap is already very high ($>99.9\ \%$) for Coulomb interaction -- a fact that contributed to the initial success of Laughlin's \emph{ansatz} as a description of the FQH effect.
However, as the power-law is made shorter ranged by increasing $p$, the overlap increases further and rapidly approaches $1$ as $p \to 4$, in agreement with the theoretical arguments made above, and in a manner compatible with the response to anisotropy plotted in Fig.\ \ref{fig:fermion_powerlaw}.

\subsection{Fermionic FQHE daughter state at $\nu=2/5$} \label{sec:fermionicdaughter}

Many different FQH fractions $\nu \neq 1/m$ have been explained by the hierarchy \cite{Haldane1983} or composite fermion (CF) \cite{Jain1989, Jain2007} pictures.
These can inform our expectation of the response of these states to anisotropy, and we briefly review them in the following.

In the hierarchy picture, introducing a number $N_{qh} = (N_e + 1) / 2p$ of quasiholes or quasielectrons into the Laughlin state at a parent filling $\nu = 1/q$ creates a daughter state of quasiparticles at electronic filling $\frac{2p}{2pq \pm 1}$.
This daughter state is decribed by a Laughlin wavefunction $\laughlin{q}$ of quasiparticles.
Starting with any daughter state, the process can be recursed to obtain a whole tree of states originating from a single parent.
The hierarchy states are FQH liquids provided the pseudopotentials for quasiparticles decay sufficiently quickly and the energy gaps are large enough.
Since the wavefunctions of daughter states are related to those of the parent states, we would expect the response to anisotropy to follow similarly.

In the CF picture, one starts from the concept of flux attachment: 
$2p$ quanta of magnetic flux are attached to each electron so that the resulting object, the composite fermion, sees an effective filling $\nu_{CF} = \frac{\nu}{1-2p\nu}$.
Then the integer quantum Hall effect of CFs, $\nu_{CF} = n \in \mathbb Z$, explains electronic fractions $\frac{n}{1+2pn}$.
The original Laughlin state corresponds to $n = 1$, while other integers form a ``Jain sequence'' that culminates in the even-denominator CFL state $\nu = \frac{1}{2p}$.
In this picture, since the entire Jain sequence is ultimately created from the same object, it is natural to associate a shape to the CF and expect it to be inherited by all fractions in the sequence.

\begin{figure}[ht!]
\centering
\includegraphics[width=\columnwidth]{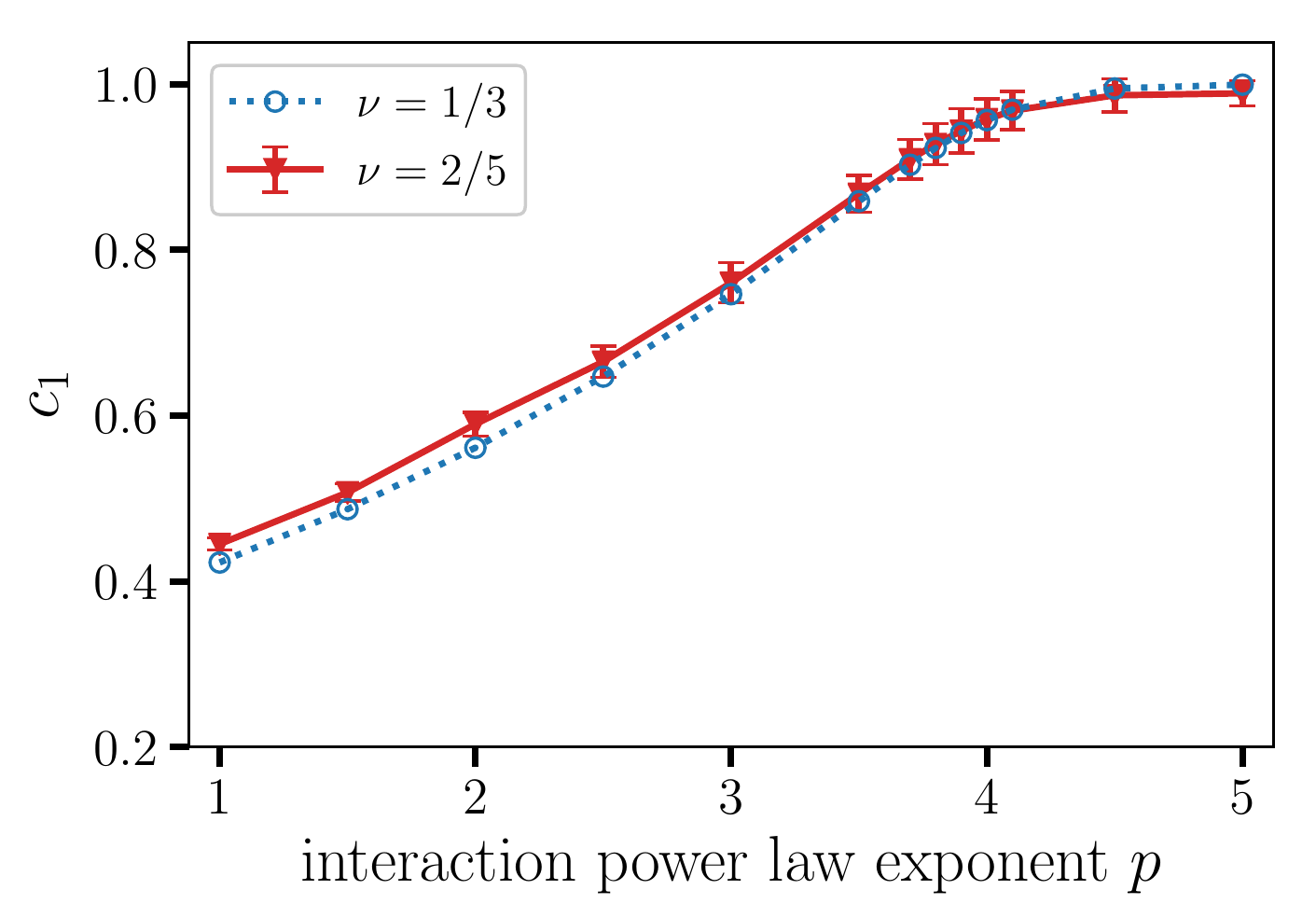}
\caption{The linear anisotropy coefficient $c_1$ for the FQH state at filling $\nu=2/5$.
The short-distance cut-off is fixed at $\Delta = 10^{-3}$.
For comparison, the curve from Fig.\ \ref{fig:fermion_powerlaw} for $\nu = 1/3$ is plotted as a dashed blue line.
}\label{fig:fermion_2_5_powerlaw}
\end{figure}

In Fig.\ \ref{fig:fermion_2_5_powerlaw}, we plot the numerical fit to linear anisotropy coefficient $c_1$ at filling $\nu = 2/5$.
This state is a daughter of the $\nu = 1/3$ state, and based on the previous discussion we expect it to show a similar response as its parent state.
In line with our expectations, we see quantitative similarities over the entire range of interaction with the plot of $c_1$ for the $\nu = 1/3$ state in Fig.\ \ref{fig:fermion_powerlaw}.
In particular we again see a critical power law exponent $p_c = 4$ above which the anisotropy is transferred completely to the FQH state.

\subsection{Composite Fermi liquid state at $\nu = 1/2$} \label{sec:CFL}

\begin{figure}[ht!]
\centering
\includegraphics[width=\columnwidth]{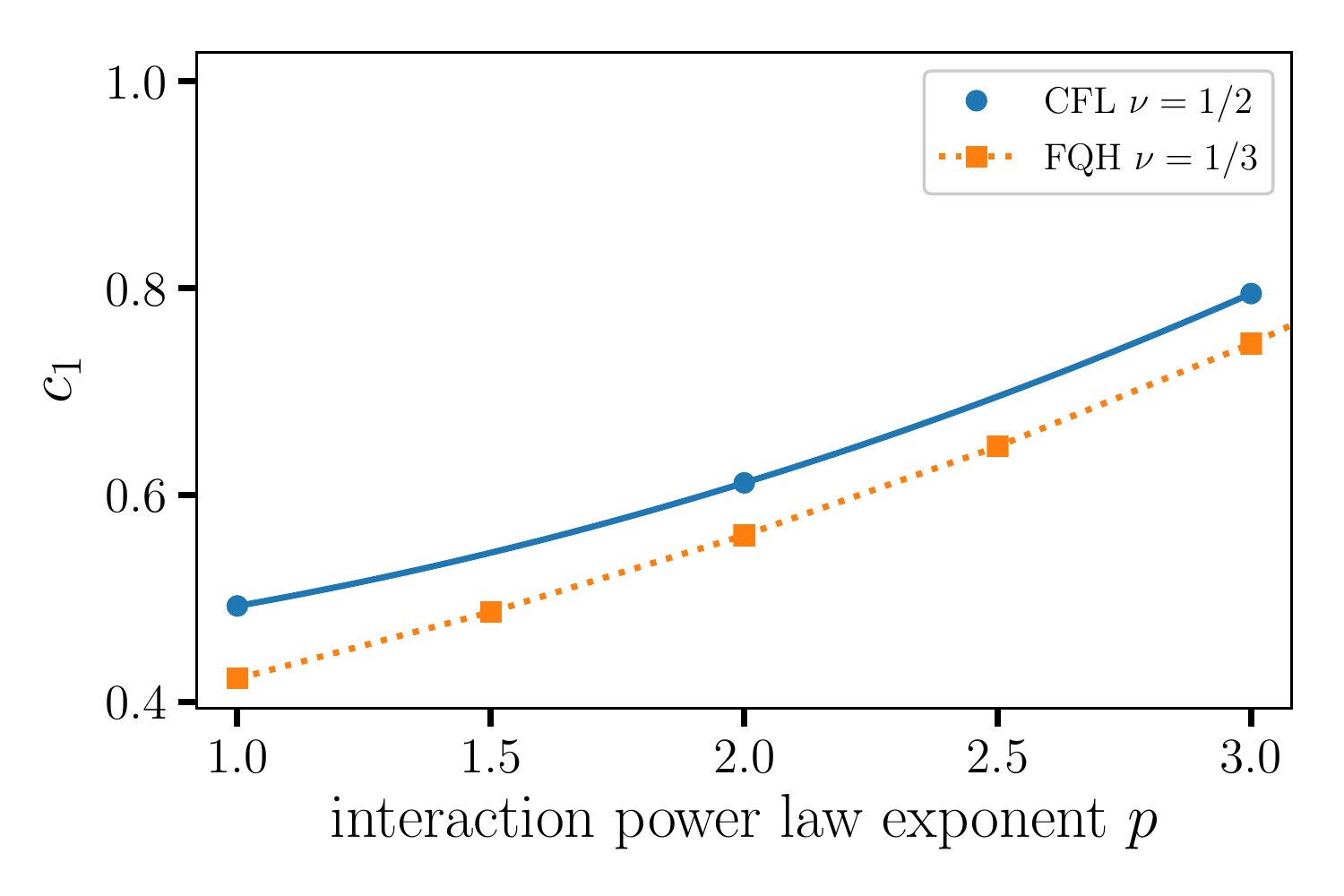}
\caption{The linear anisotropy coefficient $c_1$ for the $\nu=1/2$ CFL, for three different power laws.
For comparison, the curve from Fig.\ \ref{fig:fermion_powerlaw} for $\nu = 1/3$ is also plotted.}
\label{fig:CFL_powerlaw}
\end{figure}

In Ref.\ \cite{Ippoliti2017}, the transference of anisotropy to the CFL for Coulomb ($1/r$) and dipolar ($1/r^3$) interactions was studied.
The coefficient $c_1$ for the two cases was found to be $\simeq 0.49$ and $\simeq 0.80$.
Here, we perform calculations for the intermediate case $V(r) = 1/r^2$, using system sizes $13l_B \leq L_y \leq 24l_B$ and bond dimension $\chi = 4096$.
Following the method outlined in Sec.~\ref{sec:model}, we find a coefficient $c_1 \simeq 0.61$ (Fig.\ \ref{fig:CFL_powerlaw}).
Unlike the gapped FQH case, here each point is calculated by aggregating data over all sizes, which makes it impossible to characterize the error from the variation of $c_1$ with system size.
We instead estimate the uncertainty to be approximately $0.02$, based on the extremes of acceptable fits to the elliptical Fermi surface.

This result is consistent with our understanding that the transference of anisotropy to the CFL becomes larger as the power-law exponent $p$ increases.
%This is because as $p$ increases, the range of interaction becomes smaller and it becomes less effective at screening the effects of anisotropic band mass.
It is also interesting that the $c_1$ coefficient extracted for the CFL is consistently larger than that for the $\nu = 1/3$ FQH state and its daughter state $\nu = 2/5$:
the discrepancy is small, especially considering the difficulty of analyzing finite-size effects on the CFL results, but it is nonetheless reproduced systematically, for three distinct power laws, with similar magnitude.
The hierarchy sequence starting from the $\nu=1/3$ FQH state culminates in the CFL at $\nu=1/2$. 
It is an interesting open question whether the response to anisotropy of gapped FQH states in the sequence drifts smoothly towards the CFL value as $\nu$ is increased, or whether the response changes discontinuously as the gap closes.
The small discrepancy between $\nu = 2/5$ and $\nu = 1/3$ seen in Fig.~\ref{fig:fermion_2_5_powerlaw} seems to support the former scenario;
however, more work is needed to settle this issue.

%%%%%%%%%%%%%%%%%%%%
%%%%%%%%%%%%%%%%%%%%
%%%%%%%%%%%%%%%%%%%%
%%%%%%%%%%%%%%%%%%%%

\section{Bosonic states \label{sec:bosonic}}

The FQH can also be realized in systems of bosons with a repulsive interaction in an external magnetic field.
In this section, we repeat the analysis of fermionic FQH states of Sec.~\ref{sec:fermionic} on the analogous states for bosonic systems. 
Our results, including the singularity at power law exponent $p_c(m) = 2(m-1)$ and the identical response of parent and daughter states, should apply regardless of the statistics of the underlying constituents.
Bosonic FQH states are therefore a natural testbed for our results.

The projected LLL Hamiltonian remains the same as in Eq.\ \eqref{eq:hamiltonianLLL}.
In this case, the ground state at filling fractions $\nu = 1/m$ is a gapped incompressible liquid for \emph{even} $m$.
The ground state energy $E_{\text{gs}}$ depends only the even pseudopotentials $V_{2m}$.
The bosonic Laughlin wavefunction at filling $\nu=1/m$ with even $m$ is the exact, zero-energy, maximum density ground state of a potential whose only non-zero Haldane pseudopotentials are $V_{2k}$, $k<m/2$.
A gapped FQH phase, adiabatically connected to the Laughlin state, is stable when higher pseudopotentials $V_2, V_4, \cdots$ are turned on \cite{Wang2011, Grass2018}, as is the case for Coulomb interactions.

The same argument we formulated for anisotropy of fermions applies to bosons as well:
any power law $V(r)\sim r^{-p}$ with $p\ge 2(m-1)$ maps onto a contact interaction and gives $c_1=1$, whereas lower powers may give nontrivial response $0<c_1<1$.

Below we investigate the effects of anisotropy on bosonic FQH states at $\nu = 1/2, 1/4$ and $2/3$.

%\subsection{Parent states at $\nu = 1/2$ and $1/4$} \label{sec:bosonicparent}

\begin{figure}[ht!]
\centering
\includegraphics[width=\columnwidth]{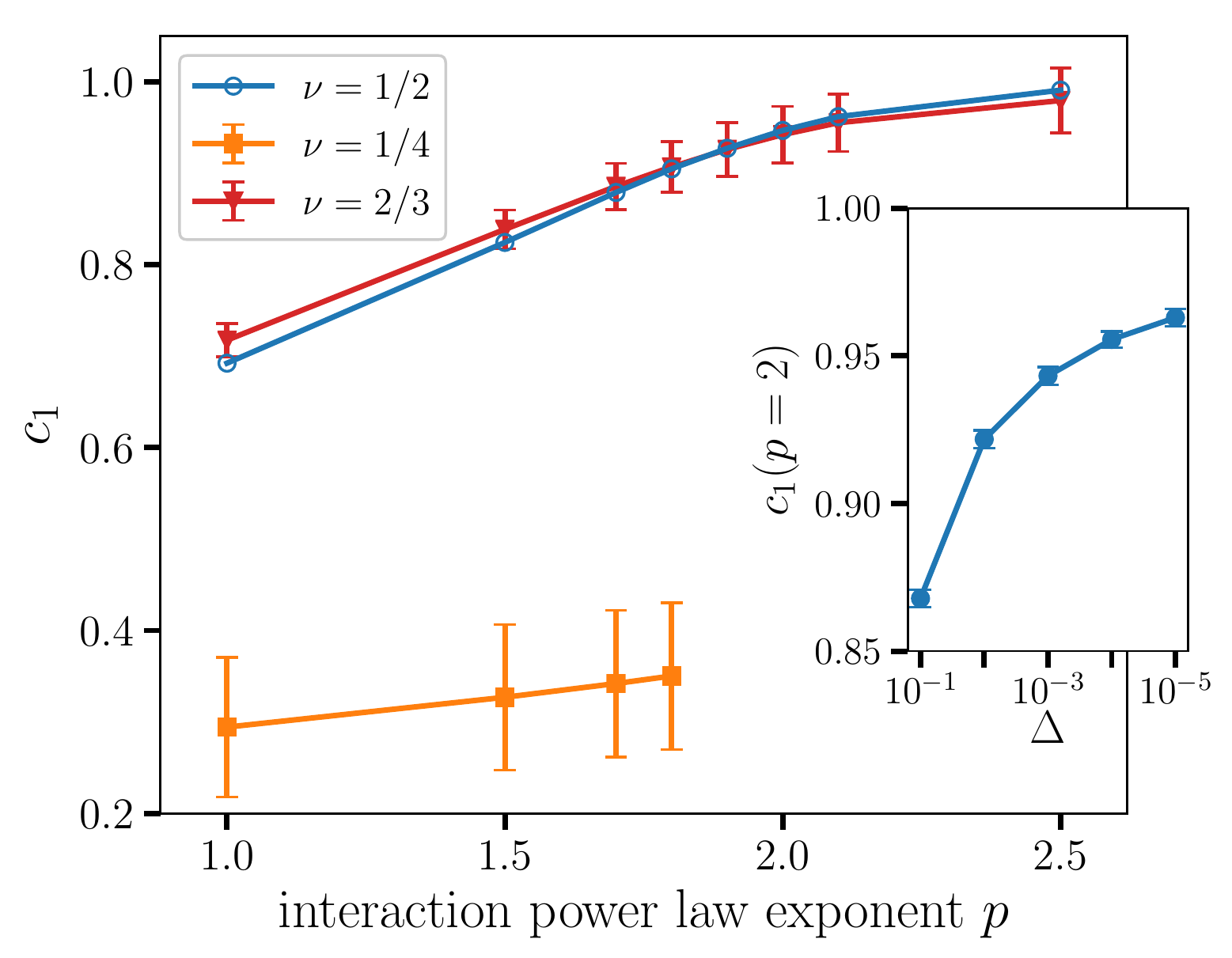}
\caption{Similar to Fig.\ \ref{fig:fermion_powerlaw}, but for bosonic parent fractions at $\nu = 1/2$ and $\nu = 1/4$, and the bosonic daughter state at $\nu = 2/3$.
The variation of linear anisotropy coefficient $c_1$ with power-law exponent $p$ is plotted.
For simulations, a cut-off $\Delta = 10^{-3}$ is used.
Points are estimated by running simulations over 5 different sizes $L_y$ and bond dimension $\chi = 2048$ to $4096$.
The inset exhibits the variation of $c_1$ with $\Delta$ for $p_c=2$ for filling $\nu = 1/2$, showing the convergence of $c_1(p_c) \to 1$ in the $\Delta \to 1$ limit.}\label{fig:boson_powerlaw}
\end{figure} 

For the bosonic FQH state at $\nu = 1/2$, with Coulomb interaction, we expect the transference of anisotropy to be partial, with $c_1 < 1$, since the critical power law $p_c = 2$.
For $\nu = 1/4$, the critical power law $p_c = 6$.

In Fig.\ \ref{fig:boson_powerlaw}, we see that the linear anisotropy coefficient $c_1 \approx 0.69$ for Coulomb interactions at $\nu = 1/2$, and rises monotonically with $p$.
There is a kink at $p=2$, which is softened by our use of a short length scale cutoff $\Delta$.
For interaction power law  exponent $p \ge 2$, $c_1 = 1$.

For the state at $\nu = 1/4$, the value of $c_1$ is smaller than that for $\nu = 1/2$.
Numerical instability limits our investigations to small power laws, but nevertheless the slow growth in $c_1(p)$ is consistent with a kink in $c_1(p)$ at a much larger power-law exponent.

The bosonic FQH state at $\nu = 2/3$ is a daughter of the $\nu = 1/2$ state.
The responses of both states to anisotropy are found to be very close to each other, within the numerical accuracy of our method.
This result parallels the one we found for fermionic daughter states in the previous section.

%\subsection{Bosonic daughter state at $\nu = 2/3$ } \label{sec:bosonicdaughter}

%%%%%%%%%%%%%%%%%%%%
%%%%%%%%%%%%%%%%%%%%
%%%%%%%%%%%%%%%%%%%%
%%%%%%%%%%%%%%%%%%%%

\section{$C_4$-symmetric distortions \label{sec:fourfold}}

In this Section we extend our discussion beyond elliptical anisotropy to consider band dispersions with discrete four-fold ($C_4$) rotational symmetry, which is often present in real band structures but does not seem to play as much of a role as band mass ($C_2$) anisotropy in shaping the FQH state.  
These type of distortions have recently gained attention in contexts ranging from the integer quantum Hall effect~\cite{Shen2015}, a field-theoretic approach to the CFL~\cite{Nguyen2017}, and the out-of-equilibrium dynamics of the FQH ``graviton''~\cite{Liu2018}.
Generalized anisotropic pseudopotentials~\cite{Yang2017} have been developed to address this and other types of distortions beyond band mass anisotropy.

Previous numerical work~\cite{Ippoliti2017b} has analyzed this problem for the CFL at filling $\nu=1/2$, where the effect of $C_4$ symmetric distortions was found to be substantially smaller than that of $C_2$ distortions (by about one order of magnitude at the level of linear response).
The distortions were measured from the shape of the CFL Fermi contour via the same method outlined in Sec.~\ref{sec:CFL}, which probes momenta $q \approx \ell_B^{-1}$.
This raises the question of long-wavelength ($q\to 0$) response in incompressible FQH states, where the quartic behavior of $S(\mb q)$ may provide a natural channel for $C_4$ symmetric distortions and thus one may expect stronger effects.

\subsection{Model and method}

We follow the method used Ref.~\cite{Ippoliti2017b}, which we review below.
We consider the dispersion
\begin{equation}
\varepsilon(k,\theta) = k^4 (1+\tanh (2\gamma) \cos(4\theta)) 
\equiv E_F \left( \frac{k}{k_F(\theta)}\right)^4 \;,
\label{eq:4folddisp}
\end{equation}
which is $C_4$-symmetric, is a polynomial in $k_x$, $k_y$, and defines a Fermi contour $k_F(\theta)$ whose overall magnitude depends on electron density ($n$), but whose shape depends only on $\gamma$, not $n$. As a result, the zero-field Fermi surface for any electron density is characterized by a fixed anisotropy
\begin{equation}
\alpha \equiv \frac{k_F(\pi/4)}{k_F(0)} = e^\gamma \;.
\label{eq:4foldfermicontour}
\end{equation}
From Eq.~\eqref{eq:4folddisp} we calculate the generalized LLL orbital by quantizing \begin{equation}
k_x \mapsto \frac{a + a^\dagger}{\sqrt{2} l_B} 
\quad
k_y \mapsto \frac{a-a^\dagger}{i\sqrt{2}l_B}
\end{equation}
and numerically finding the ground state of the resulting sparse Hamiltonian.
The Landau levevel mixing coefficients in the expansion of the ground state $\ket{\tilde{0}}$ in the basis of isotropic Landau levels $\{ \ket{N}: N \geq 0 \}$, 
\begin{equation}
\ket{\tilde{0}} \equiv \sum_N u_N \ket{N} \;,
\end{equation}
are such that $u_N \neq 0$ only for $N = 0, 4, 8\dots$ because of the $C_4$ symmetry.
These coefficients can be used to calculate the anisotropic form factor $F_{\tilde{0} \tilde{0}}(\mb q)$ as a linear combination of isotropic ones, which are known analytically.
The resulting interaction is then used to build the matrix product operator Hamiltonian for the iDMRG method.

In the 2D thermodynamic limit, the system has $C_4$ symmetry, so the guiding center structure factor of the many-body FQH ground state, $S(q)$, must be of the form
\begin{equation}
S(q,\theta) = e^{2D} \left(1+\tanh(2\sigma)\cos(4\theta)\right) q^4 \;.
\end{equation}
$D$ is even under $C_4$, while $\sigma$ and $\gamma$ are odd.
This directly generalizes the definitions of $\gamma$, $D$ and $\sigma$ used previously for the case of band mass anisotropy.
It remains true, in particular, that $\aQH = e^\sigma$, if one defines $\aQH$ as the anisotropy of equal-value contours of $S(q)$ at $q\ll l_B^{-1}$, in analogy to Eq.~\eqref{eq:4foldfermicontour}.
Letting 
\begin{equation}
\lambda = \lim_{q_x\to 0} S(q_x,0)/q_x^4 \;,
\end{equation}
we have
\begin{equation}
\sigma = \frac{1}{4} \ln \frac{\lambda(\gamma)}{\lambda(-\gamma)}  \;, \quad
D =  \frac{1}{2} \ln \left( \frac{\lambda(\gamma) + \lambda(-\gamma)}{2} \right) \;.
\end{equation}
In the following we focus on fermionic states at fillings $\nu = 1/3$ and $1/5$ with Coulomb interaction.

\subsection{Laughlin state, $\nu = 1/3$}

\begin{figure}
\centering
\includegraphics[width=\columnwidth]{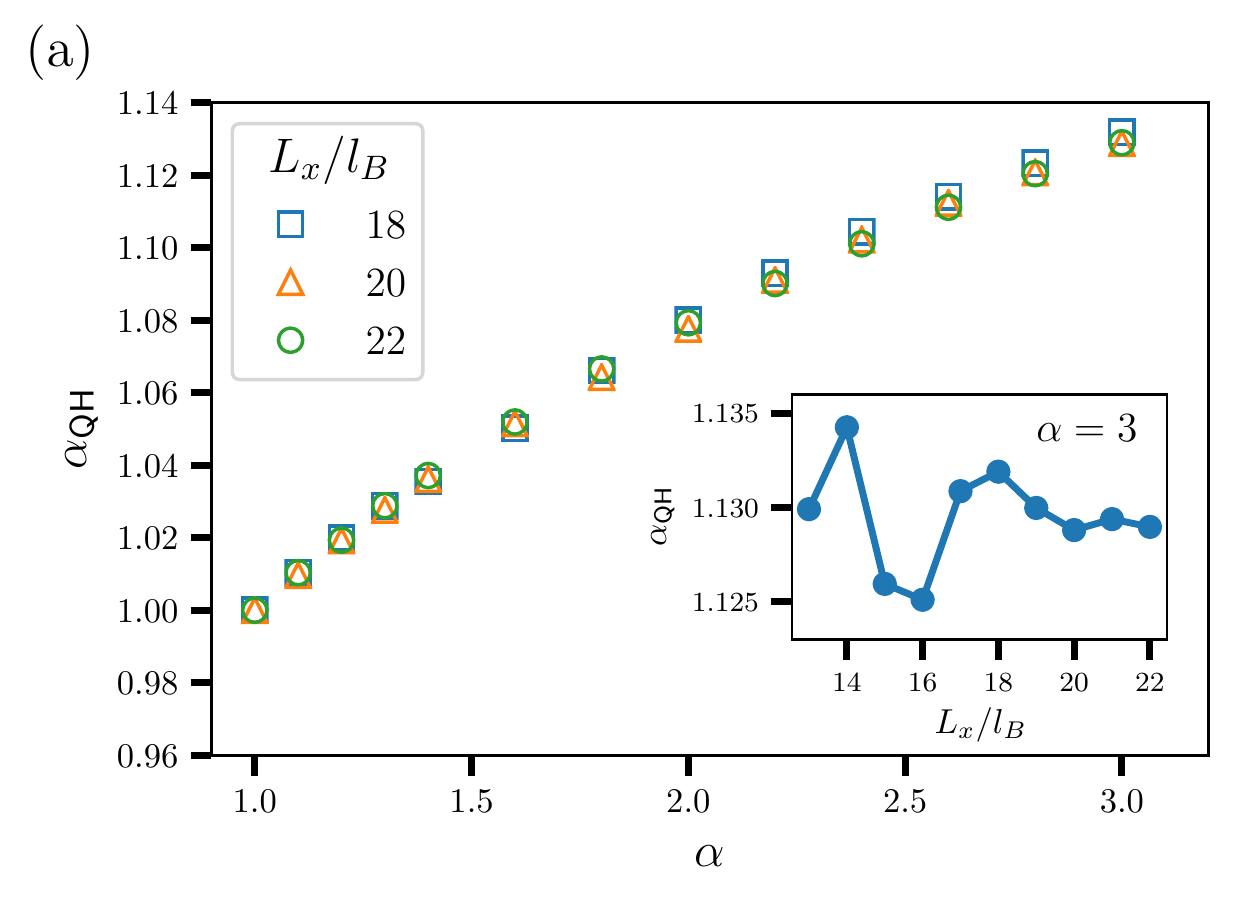}
\includegraphics[width=\columnwidth]{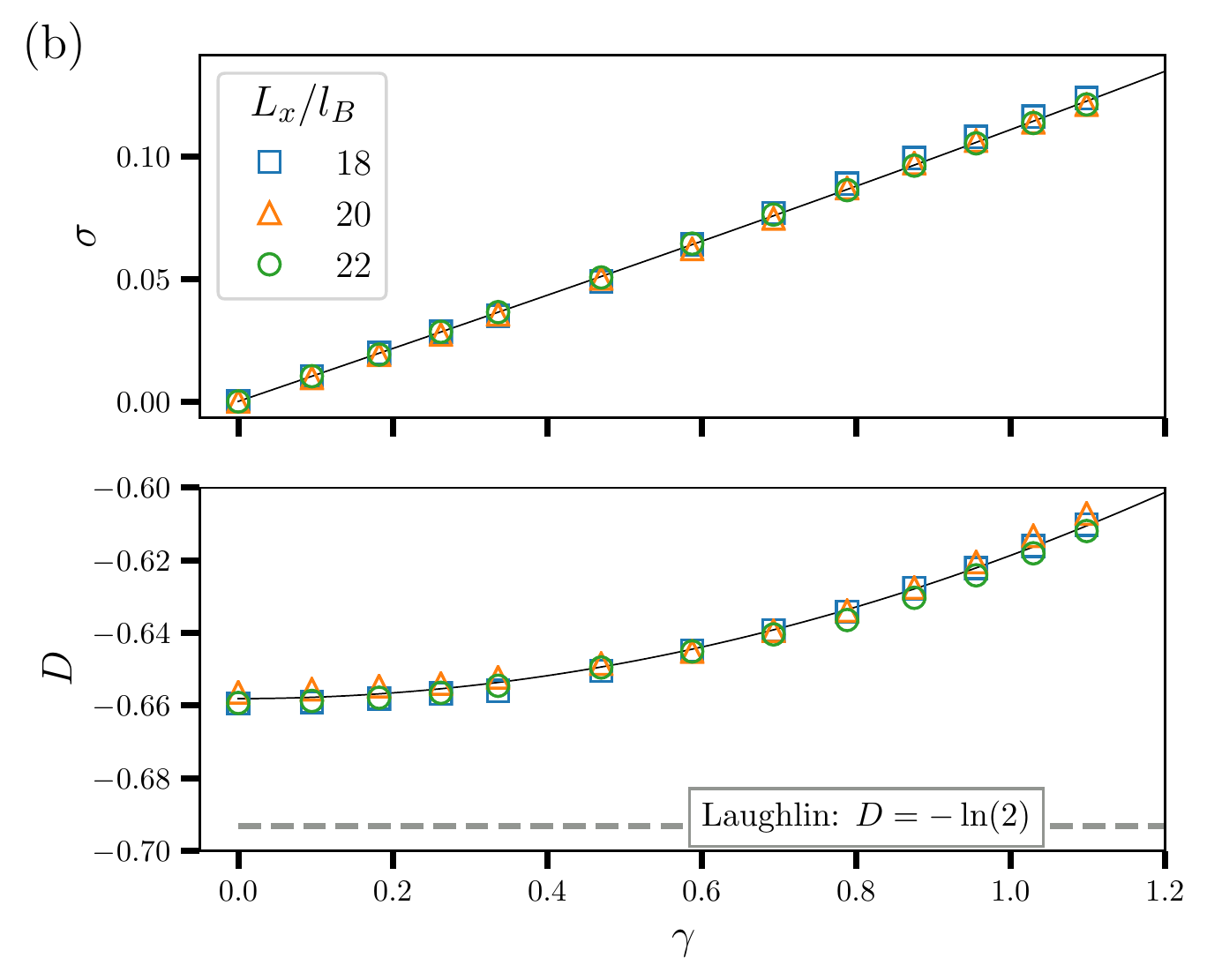}
\caption{Numerical results for $\nu = 1/3$ with Coulomb interaction,
obtained with DMRG bond dimension $\chi = 1000$.
(a) Anisotropy of the FQH state $\aQH = e^\sigma$ as a function of band anisotropy $\alpha  = e^\gamma$. Inset: finite-size oscillations of $\aQH$ at fixed $\alpha=3$.
(b) Logarithmic parameters $\sigma$ and $D$ as a function of $\gamma$. 
Finite size effects are very small. 
Continuous lines correspond to fits of the data at $L_x = 21l_B$ to fixed-parity polynomials of the form $\sigma = c_1 \gamma + c_3 \gamma^3$ and $D = c_0 + c_2\gamma^2$.
Results for data averaged over all sizes are similar.
The dashed line shows the lower bound $D \geq -\frac{1}{2} \ln(2)$, achieved by the Laughlin state $\laughlin{3}$ (with pure $V_1$ interaction), for comparison.
\label{fig:4fold13}}
\end{figure}

For the $\nu = 1/3$ state, we obtain the results shown in Fig.~\ref{fig:4fold13}.
The functions $\sigma(\gamma)$ and $D(\gamma)$ are approximately given by 
\begin{equation}
\sigma(\gamma) \simeq 0.11 \gamma  \; ,
\quad
D(\gamma) \simeq -0.66 + 0.04 \gamma^2  \;,
\label{eq:4fold_13_result}
\end{equation}
where the cubic term in $\sigma$ is found to be compatible with zero: $c_3 = 0 \pm 0.01$.
The linear term in the response $\sigma$ is significantly larger than what was found in the CFL~\cite{Ippoliti2017} ($c_1 \simeq 0.06$ in this paper's notation),
but still only a quarter (i.e.\ much smaller than) the magnitude of the response to $C_2$ band mass anisotropy, $c_1\simeq 0.43$.
Another striking difference with respect to the elliptical case is that the isotropic dilation, parametrized by $D$, is not constant.
It is, on the contrary, of comparable magnitude as the distortion $\sigma$ itself.
The quadratic coefficient $c_2\simeq 0.04$ is not compatible with zero, as is clear from Fig.~\ref{fig:4fold13}.

\subsection{Laughlin state, $\nu = 1/5$}

The $\nu = 1/5$ state, as discussed in Sec.~\ref{sec:fermionicparent}, has significantly weaker response than the $\nu = 1/3$ state.
In Ref.~\cite{Ippoliti2018} this was attributed to a general feature of flux attachment:
the single-particle orbitals attached to each electron are most sensitive to anisotropy near the core, and get progressively closer to circular as one moves outwards;
therefore attaching more fluxes (i.e. lowering the filling) gives rise to \emph{less anisotropic} FQH states.

\begin{figure}
\centering
\includegraphics[width=\columnwidth]{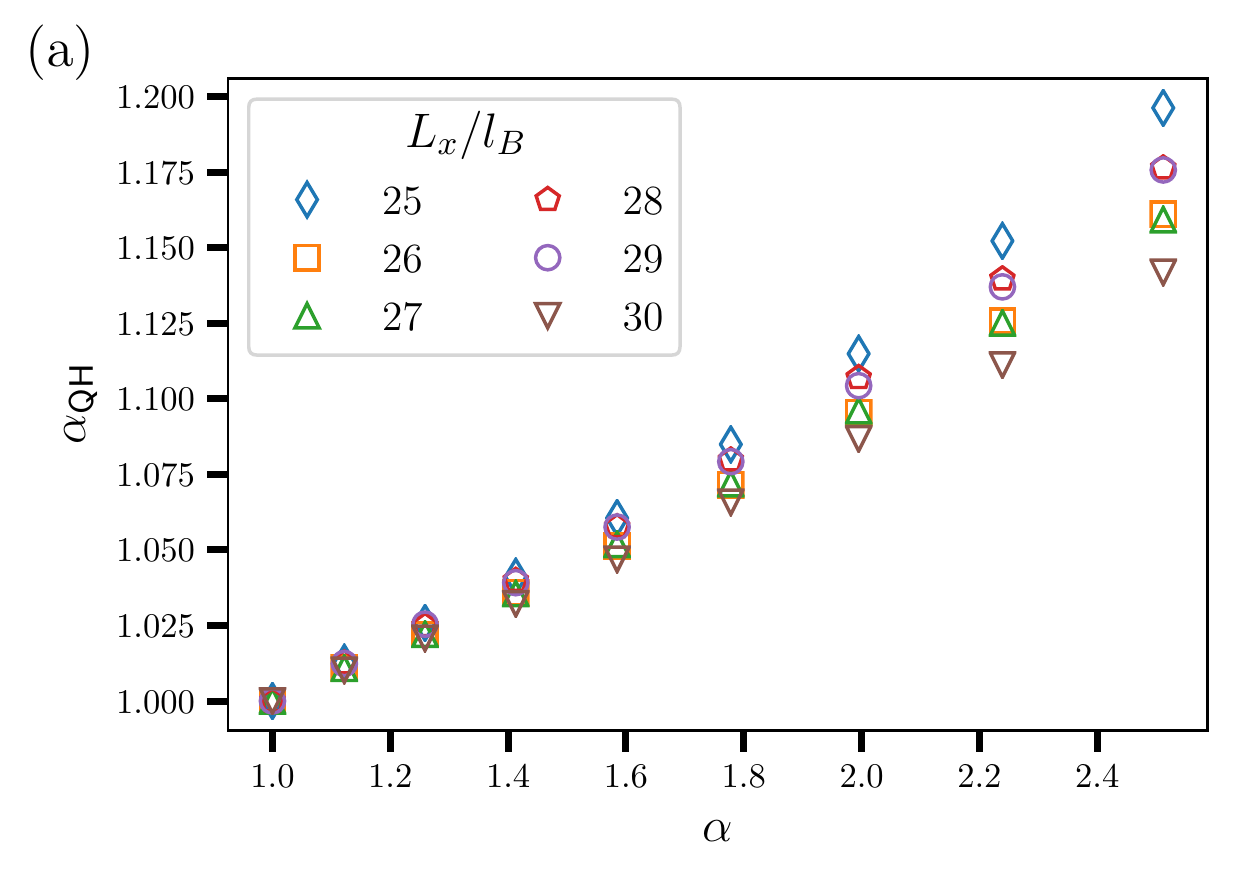}
\includegraphics[width=\columnwidth]{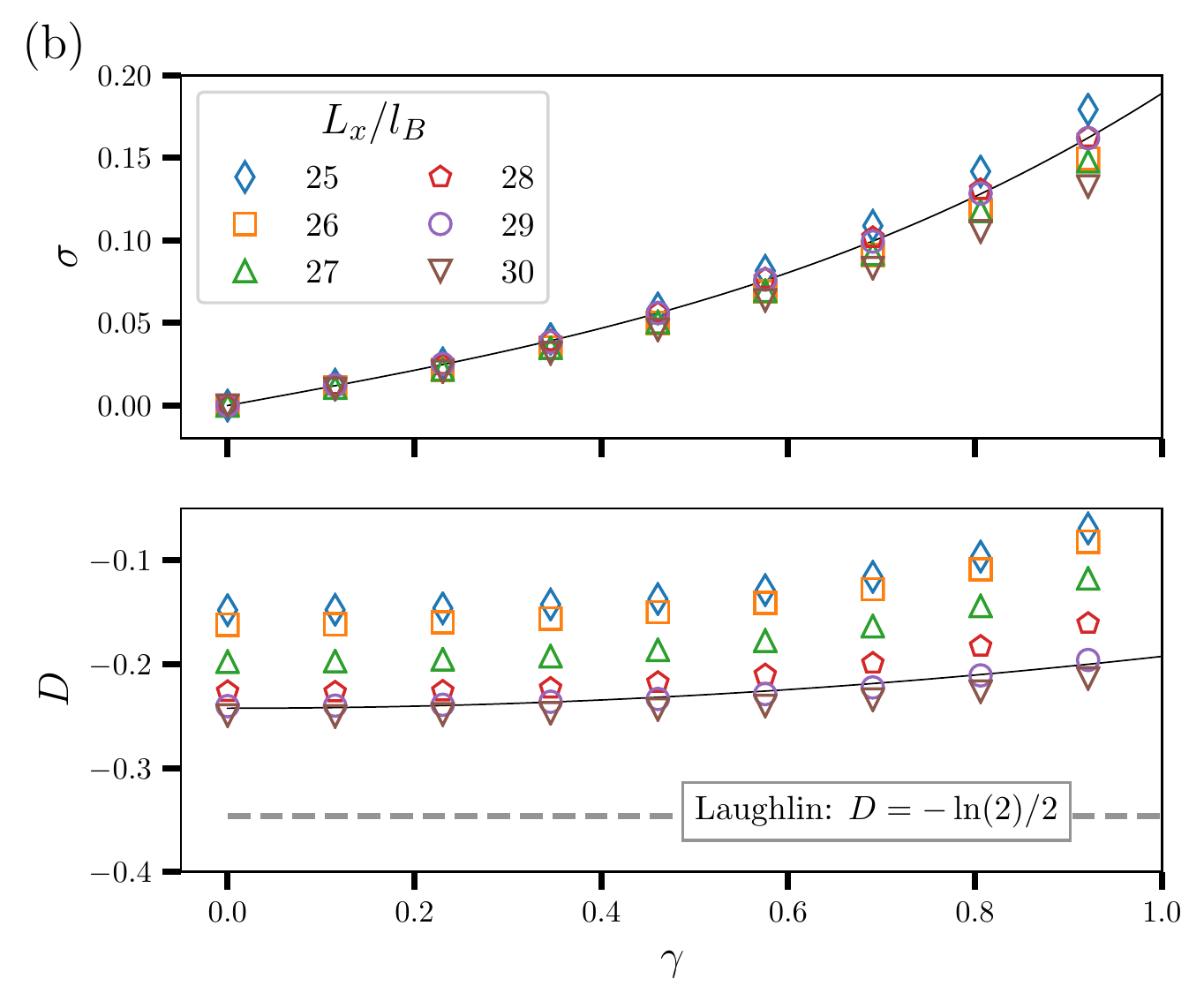}
\caption{Numerical results for $\nu = 1/5$ with Coulomb interactions, obtained with DMRG bond dimension $\chi = 2000$.
(a) Anisotropy of the FQH state $\aQH = e^\sigma$ as a function of band anisotropy $\alpha  = e^\gamma$. 
(b) Logarithmic parameters $\sigma$ and $D$ as a function of $\gamma$. 
Finite size effects are stronger in this case, despite the larger sizes considered (up to $L_x = 30 l_B$). 
Continuous lines correspond to fits of the data at $L_x = 29 l_B$ to the same polynomial forms as in Fig.~\ref{fig:4fold13}.
The drift of $D$ with size must stop before saturating the lower bound $D \geq -\ln(2)$ attained by $\laughlin{5}$.
\label{fig:4fold15}}
\end{figure}

The results for $\nu = 1/5$, shown in Fig.~\ref{fig:4fold15}, are therefore surprising. 
We find
\begin{equation}
\sigma(\gamma) \simeq 0.10 \gamma  + 0.08 \gamma^3 \;,
\quad
D(\gamma) \simeq -0.24 + 0.05 \gamma^2  \; .
\label{eq:4fold_15_result}
\end{equation}
The magnitude of the response $\sigma$ is \emph{not} smaller than that of the $\nu = 1/3$ state: the linear coefficients $c_1$ are compatible within finite-size uncertainty, while the cubic coefficient $c_3$ is significantly larger in this case. 
We also find that the isotropic rescaling $D$, despite much stronger finite-size effects, appears to be consistent with that of the $\nu = 1/3$ state, up to an expected shift in the constant term:
for the Laughlin state $\laughlin{m}$ one has $D(0) = \frac{1}{2} \ln \frac{m-1}{8}$; this value sets a lower bound for realistic interactions, e.g. Coulomb\cite{Haldane2009}, hence the offset in going from $m=3$ to $m=5$.

The response $\sigma$ in Eq.~\eqref{eq:4fold_15_result} is remarkable because it also signals a qualitatively distinct behavior of the $\nu = 1/5$ state.
For $\nu = 1/3$, the anisotropy $\aQH$ is found to be a concave function of $\alpha$ (close to $\aQH = \alpha^{0.11}$), which may be related to the apparent saturation of composite fermion anisotropy $\alpha_{\sf CF}$ observed in Ref.~\cite{Ippoliti2017b}.
On the contrary, for $\nu = 1/5$ we find that $\aQH$ is a \emph{convex} function of $\alpha$, growing super-linearly in the interval $1\leq \alpha \lesssim 3$ that we investigated numerically, and suggesting that much stronger distortions may be possible at larger $\alpha$.

This overall stronger response is in striking contrast to the results for band mass anisotropy and their interpretation given in Ref.~\cite{Ippoliti2018}.
Secondly, the isotropic rescaling $D$, despite much stronger finite-size effects, appears to be consistent with that of the $\nu = 1/3$ state, up to an expected shift in the constant term. 
Since the Laughlin state $\laughlin{m}$ has $D(0) = \frac{1}{2} \ln \frac{m-1}{8}$, in going from $m=3$ to $m=5$ one expects a shift of $\frac{1}{2} \ln{2} \approx  0.34$ even for Coulomb interaction. 

This counterintuitive result may be a special feature of the $\nu = 1/5$ state.
At filling $\nu = 1/m$, one can write a deformed Laughlin wavefunction with exact $C_{m-1}$ symmetry as follows\cite{Ciftja2004}:
\begin{align}
\Psi_m(\{z\})
& = e^{-\frac{1}{4} \sum_i |z_i|^2} \prod_{i<j} (z_i-z_j)  \nonumber \\
& \qquad \times \prod_{\mu = 0}^{m-1} \left(z_i -z_j - \eta e^{2\pi i \mu/m} \right) \;,
\label{eq:model_nfold}
\end{align}
where $\eta \equiv |\eta| e^{i\phi} $ is a parameter controlling the magnitude ($|\eta|$) and orientation ($\phi$) of the distortion.
In other words it is possible to split the $m$-fold zero into a single zero (necessary for fermionic antisymmetry) and $m-1$ zeros arranged on the vertices of a regular polygon, which has $C_{m-1}$ discrete rotational symmetry.
One can thus construct a $C_4$-symmetric Laughlin-like state for $\nu=1/5$, but not $\nu=1/3$. 

On the other hand, a $C_N$ symmetric state can be constructed for all even $N<m$:
for example, by modifying the prescription in Eq.~\eqref{eq:model_nfold} to 
\begin{align}
\Psi_m(\{z\})
& = e^{-\frac{1}{4} \sum_i |z_i|^2} \prod_{i<j} (z_i-z_j)^{m-N} \nonumber \\
& \qquad \times \prod_{\mu = 1}^N \left ( z_i-z_j+\eta e^{2\pi i \mu/N} \right) \;.
\end{align}
This suggests that the Laughlin state $\laughlin{m}$ may have a natural way of responding to distortions with $C_N$ symmetry for $N \leq m-1$, but \emph{not} for larger $N$. 
This would explain the weaker response of $\nu = 1/3$ to $C_4$ distortions relative to $\nu = 1/5$.

Unfortunately this conjecture is hard to test beyond this level, for two reasons:
(i) the difficulty in studying the next Laughlin state, $\laughlin{7}$ at $\nu =1/7$, 
and (ii) the difficulty in identifying and calculating observables which can display $C_6$-symmetric distortions.
At filling $\nu = 1/7$ Coulomb interactions favor a Wigner crystal over the incompressible FQH state~\cite{Lam1984}, and while the Laughlin state can be engineered numerically for suitably short-ranged interactions, finite-size effects are bound to be much worse than for the $\nu = 1/3$ and $1/5$ states.
As for signatures of $C_6$-symmetric anisotropy,
any anisotropic terms in $S(q)$ at small $q$ would occur at $\mathcal O(q^6)$ and would thus be drowned out by the leading isotropic term $q^4$. 
One would need to identify higher-order correlators with a leading $q^6$ behavior, which would also be considerably harder to calculate numerically. 

These issues, while interesting and still largely unexplored, go beyond the scope of the present work and we leave them to future investigations.

%%%%%%%%%%%%%%%%%%%%
%%%%%%%%%%%%%%%%%%%%
%%%%%%%%%%%%%%%%%%%%
%%%%%%%%%%%%%%%%%%%%

\section{Discussion \label{sec:discussion}}

We have systematically studied the response of fractional quantum Hall states to geometric distortions for a wide variety of interaction potentials and LLL filling fractions, for both fermionic and bosonic particles. 
We considered isotropic interactions and introduced anisotropy by means of the single-particle dispersion.
The anisotropy of the FQH ground state then results from a competition between the shape of interactions and that of single-particle orbitals.

We have performed infinite density matrix renormalization group (iDMRG) simulations of the problem and extracted the anisotropy of the FQH ground state from its static guiding center structure factor, in particular, from its long-wavelength limit. 
This approach relies on the ability to accurately probe very long wavelength in one direction, which is a unique strength of the infinite DMRG method.

Generically, for power-law interactions $V(r)\sim r^{-p}$, we found confirmation to the intuition that larger values of $p$ correspond to ``shorter-range'' interactions and thus are less effective at washing out the anisotropy of single-particle orbitals.
Even though all power law interactions are strictly speaking long-range, we make the above intuition more rigorous by considering their pseudopotential decomposition.
Doing so reveals singularities at special values of the power law exponent $p$, beyond which interactions effectively transition from long range to contact.
When this happens, only one geometry is left in the problem, and the FQH fluid simply inherits the same anisotropy as the single-particle orbitals.

In particular, we found that for filling $1/m$ (with $m$ even for fermions and odd for bosons) this transition occurs at $p = 2(m-1)$.
This has several interesting implications.
For one, it supports the idea, presented in Ref.~\cite{Ippoliti2018}, that Laughlin states in the presence of band mass anisotropy should get \emph{less anisotropic} with decreasing filling.
Bosonic $\nu = 1/2$ achieves maximal anisotropy at $p=2$, followed by fermionic $\nu=1/3$ at $p=4$, etc;
low-filling states thus achieve maximal anisotropy only for very large $p$.
Combined with our numerical data for $1\leq p \lesssim 4$, this strongly suggests that the transference of anisotropy to this type of FQH states is monotonically decreasing with $m$ for generic interactions.

One more consequence of these results is the prediction that certain FQH states with particular kinds of interactions should have a trivial geometric degree of freedom; i.e., that their intrinsic metric should be completely decided by single-particle physics. 
This conclusion applies whenever $p>2(m-1)$, which applies, in particular, to the bosonic $\nu=1/2$ state with dipolar ($p=3$) interaction. 
Interestingly, this state could be realized using ultracold polar molecules in an optical lattice with synthetic gauge potentials~\cite{Yao2013}.

We also find that ``daughter states'' derived from the Laughlin states via the hierarchy or composite fermion pictures show the same response to band mass anisotropy as their parent state, supporting the idea that parent and daughter states alike are ultimately created from the same anisotropic object. 
This was already observed for Coulomb interactions, but here we find the same result across a range of power law interactions, supporting the universality of this conclusion.
Intriguingly, the non-analyticity at $p=2(m-1)$ is found in the daughter states as well as the parent states. This does not follow trivially from the pseudopotential decomposition.

One question that remains open is that of the relationship between the $\nu=1/2$ fermionic CFL and the Jain sequence that emanates from it and culminates at $\nu = 1/3$.
Being made of the same composite fermion building blocks, we would expect the fermionic $\nu = 1/2$ and $\nu=1/3$ to have the same response. 
However, the CFL appears to have slightly stronger response to anisotropy than the $\nu = 1/3$ FQH state (though much weaker response than the bosonic $\nu = 1/2$ FQH state).
This may point to a slow drift of the response along the Jain sequence, which we fail to resolve numerically; or it may be a singular feature of the gapless CFL state that sets it apart from the gapped fractions in the sequence.
It is worth pointing out that the definition of anisotropy for FQH states relies on their incompressibility (the quartic behavior of $S(q)$ near $q=0$), while for the CFL it is based on the Fermi contour, at $q\approx \ell_B^{-1}$; the small discrepancy could be a consequence of this different definition also. 
Finally, it may also arise from finite-size effects, which are better controlled in gapped FQH states than they are in the CFL.
Consequently, the nature and explanation of this discrepancy remains an open issue.

Finally, we have extended our investigation to band distortions with discrete four-fold ($C_4$) rotational symmetry. 
There we have found surprisingly that trend of decreasing anisotropy with decreasing filling is inverted, at least in going from $\nu = 1/3$ to $\nu = 1/5$. 
We conjecture that this exception to the trend may come from the ability of Laughlin states with $\nu = 1/m$ to naturally accommodate $C_N$-symmetric distortions by displacing some of the zeros in their wavefunction in a pattern with the appropriate symmetry. 
For fermionic states, this is possible only for $N\leq m-1$, as one zero (for each pair of electrons) is fixed by antisymmetry.
In particular, for $C_4$ symmetry, this distortion is admissible at $\nu = 1/5$ but not at $\nu = 1/3$.
Developing $C_4$-symmetric distortions may thus be more energetically costly at $\nu=1/3$; as a result the ground state may be closer to the isotropic Laughlin state. 
This explanation, if correct, implies that a similar result should hold for $C_{N}$-symmetric distortions at filling $\nu = {1}/{(N+1)}$, e.g. for $C_6$ and $\nu = 1/7$. 
However this would be rather challenging to probe numerically with the method used here, and is left as another direction for future work.

\acknowledgments

The iDMRG calculations were carried out using libraries developed by Roger Mong, Michael Zaletel and the TenPy collaboration.
This work was supported by DOE BES grant DE-SC0002140.

\begin{comment}
\appendix
A triangular Wigner crystal with lattice constant $d$ at filling $\nu$ has the relation $d = \sqrt{\frac{8 \pi}{\sqrt{3} \nu}} l_B = \frac{3.809}{\sqrt{\nu}} l_B$.
\end{comment}

%\clearpage

\bibliography{iDMRG_papers}

\end{document}